\documentstyle[twocolumn,aps,prl]{revtex}

\begin{document}
\draft
\title{Conformal, Subconformal and Spectral Universality in
Incommensurate Spin Chains}
\author{ Jukka A. Ketoja}
\address{ Department of Theoretical Physics,\\ P.O. Box 9,
FIN-00014 University of Helsinki, Finland}
 \author{ Indubala I. Satija\cite{email} }
\address{
 Department of Physics and\\
 Institute of Computational Sciences and Informatics,\\
 George Mason University,\\
 Fairfax, VA 22030}
\date{\today}
\maketitle
\begin{abstract}

A renormalization scheme is developed to study an anisotropic
quantum XY spin chain in a quasiperiodic transverse field.
The critical phase of the quasi-particle excitations of the model
with fractal wave functions exists in a finite parameter interval and is
sandwiched between the extended and localized phases.
The scaling properties of the critical phase fall into
four distinct universality classes referred as
spectral, subconformal, conformal and Harper. The spectral and
conformal classes respectively describe
the onsets of extended to critical and critical to localized
transitions while the subconformal
class describes the part of the phase diagram sandwiched between
the conformal and spectral transitions. The Harper universality class describes
the isotropic limit of the XY model.
A decimation scheme is developed to compute the infinite sets
of universal scaling ratios characterizing the wave functions in the four
universality classes.
The renormalization flow equations exhibit a limit
cycle at the band center and at the band edges providing
a new method for determining these energies with extremely high precision.
\end{abstract}

\narrowtext
\section{Introduction}
In recent years, renormalization group (RG) techniques have been
the dominant theme
in describing the scaling properties at the onset
of phase transitions. These studies include the thermodynamical phase
transitions, driven by thermal fluctuations,
as the temperature is varied as well as the transitions which occur when
a control parameter is varied. The later include the transitions to chaos
in nonlinear dynamical systems and the phase
transitions in quantum
systems at zero temperature such as the metal-insulator transitions, driven by
quantum fluctuations.

One-dimensional quantum spin chains provide very simple and interesting
models that exhibit phase transitions at zero temperature
as a function of anisotropy
or magnetic field.\cite{Lieb} In some cases, the $T=0$ phase transitions of the
quantum spin chain map onto the phase transitions of a 2D classical model
in which the variable parameter is the temperature. These systems are
also of experimental interest as various
physical realizations of quantum spin chains have been found in quasi-1D
magnetic systems.\cite{Steiner}
The Ising and the anisotropic XY model in a $constant$ transverse field
provide the simplest class of quantum models\cite{Lieb} (that can
be solved exactly) that exhibit the second
order magnetic phase transition at a critical value of the
magnetic field.
The anisotropic model
belongs to the universality class of the Ising model, i.e. the critical
exponents at the onset of transition are the
same as those of the Ising model\cite{Lieb}.
Therefore, such anisotropic models with a uniaxial anisotropy are
usually referred as Ising like.

Motivated by the discovery of a quasi-crystal and the fabrication
of magnetic superlattices,
quantum spin chains in a quasiperiodic (QP) transverse
field have been investigated extensively.
Initial studies were confined to
the case where the transverse magnetic field was inhomogenious
with binary Fibonacci distribution.\cite{DS} The quasiparticle excitation
spectrum turned out to be a fractal for all parameter values.
These studies were followed by
other ones in which the magnetic field was
oscillatory with the period incommensurate with respect to the period
of the spin chain\cite{BR}. Detailed numerical work showed that
the quasiparticle excitations of the anisotropic quantum spin chain
could be extended (E), localized
(L) or critical (C) resulting in E-C and C-L transitions as the
strength of the magnetic field was varied. These three different phases
were respectively characterized by continuous, point like and singular
continuous spectrum. It should be noted that
in the isotropic
limit the eigenvalue equation for the quasiparticle excitations
of the model reduce to the the famous Harper equation exhibiting
both extended and localized phases and hence the metal-insulator transition.
\cite{Harper} The existence of a self-similar butterfly
spectrum at the onset
of transition aroused a great deal of interest in this model.\cite{Hofstadter}
The Harper equation has also been of  interest due to the fact that
it describes the two-dimensional electron gas in a magnetic
field.\cite{Sok} Furthermore, this model has attracted
 the attention
of mathematicians due to the small denominator problems.\cite{KAM}

The anisotropic XY model describes a perturbed Harper model and therefore
provides a new class of quasiperiodic models which splits the single E-L
transition of the Harper model into two transitions,
namely E-C and C-L transitions.
The existence of  a $fat$ C phase sandwiched between E and L
phases provides a new paradigm for the metal-insulator transition in
one-dimensional
incommensurate systems. Analogous to the
Harper model, in the anisotropic model both the E-C and C-L
transitions are global in the sense that all quantum states are
E, C or L simultaneously.  Fig. 1 shows the spectral phase diagram of
the model. The line in the phase diagram
corresponding to the onset of localization also corresponds to the
onset of magnetic phase transition from ordered to disordered phase
providing an interesting example of a spectral and magnetic interplay.
This line will be referred as the conformal line as along
this line, the gap in the excitation spectrum vanishes and the model is
conformally invariant.\cite{Hankel} The
E-C transition in the quasiparticle wave functions does not affect
 the magnetic
properties and hence the corresponding line will be referred as the spectral
line. The fat C region
sandwiched between the the spectral and conformal line will be called
subconformal. The conformal and spectral lines intersect at the Harper
limit.

In our previous short paper\cite{KS} , we described a new decimation scheme
to study nearest-neighbor (n.n.) tight binding models (TBM) and
applied it to QP Ising chains exhibiting the C-L transition.
The proposed scheme confirmed the existence of
a fat C phase in the Ising limit.
It also demonstrated the fact that the fat C phase was described
by two distinct fixed points which respectively described the
C-L transition at the conformally invariant point and the subconformal
C phase. Since the decimation scheme was valid only for n.n. TBM's,
it described only the Ising limit of the general anisotropic model
containing also a next-nearest-neighbor (n.n.n.) term.
Therefore, these studies did not
provide a full characterization of the C phase. In particular, the
universality class for the E-C transition and for the conformal line
(see Fig. 1) remained an open question.

In this paper we generalize our method to study the anisotropic model
which includes Ising as a special case.
The scalar decimation equations of our previous studies are
replaced by vector equations. We show that the C phase is
described by four distinct universality classes.

In section 2, we describe the model and the corresponding tight
binding form. In section 3, we discuss the decimation scheme based
on the Fibonacci lattice sites. Sections 4 and 5 respectively deal with
the qualitative as well as quantitative characterization of universal
features of the phase diagram.
In section 6, we show how the decimation scheme can be
generalized for an arbitrary path in the Farey tree.
In section
7, we discuss the RG equations.
Section 8 summarizes the various universality classes, and we end with
some conclusions in Section 9.

\section{ Spin Hamiltonian and the Tight Binding Equation }
The anisotropic XY model in a transverse field is given by
\begin{equation}
H = -\sum_i [\frac{1}{2}(J_x \sigma_i^x \sigma_{i+1}^x+J_y
 \sigma_i^y\sigma_{i+1}^y)+h_i\sigma_i^z]
\end{equation}
Here $\sigma$ are the Pauli matrices describing the spin $1/2$. $J_x$
and $J_y$ describe ferromagnetic exchange
interactions.
In our discussion,
unless otherwise stated, $h_i$ is an inhomogenious modulating field
of periodicity $\sigma$ which is incommensurate with the periodicity
of the spin chain:
\begin{equation}
h_i = \lambda cos ( 2 \pi( i \sigma + \phi ) )
\end{equation}
We choose the parameter $\sigma$ to be equal to the inverse golden mean
$\sigma = ( \sqrt5 -1)/2$. $\phi$ is a constant phase factor.

Using the Jordan-Wigner transformation, the model can be transformed
into that of a free spinless fermion Hamiltonian
\begin{eqnarray}
H&=&-\sum [\frac{1}{2}(J_x+J_y) c^{\dag}_ic_{i+1} +
\frac{1}{2}(J_x-J_y) c^{\dag}_i c^{\dag}_{i+1}+
\nonumber\\
&+& h_i c^{\dag}_ic_i  + h.c. ],
\end{eqnarray}
We see that the lack of the $O(2)$ symmetry in the spin Hamiltonian
results in a lack of the $U(1)$ symmetry ($e^{i\theta} c_i =
Uc_iU^\dagger$) in the corresponding fermion Hamiltonian. The
system can be diagonalized using the methods described by Lieb et al.
\cite{Lieb}. The resulting
coupled eigenvalue equations for the quasi-particle excitations
are\cite{Lieb}
\begin{eqnarray}
J_y \psi_{i+1}+J_x \psi_{i-1}+2h_i \psi_i &=& E \eta_i\\
J_x \eta_{i+1}+J_y \eta_{i-1}+2h_i \eta_i &=& E \psi_i
\end{eqnarray}
The above two coupled equations with n.n. terms can be combined into a single
equation with n.n.n. interaction:
\begin{eqnarray}
E^2 \psi_i &=&J_x J_y(\psi_{i-2}+\psi_{i+2}) \nonumber\\
&+& 2(J_y h_{i-1}+J_x h_i)\psi_{i-1}+
2(J_y h_i+J_x h_{i+1})\psi_{i+1}\nonumber\\
&+& (J_x^2+J_y^2+4h_i^2)\psi_i ,
\label{SPINTBM}
\end{eqnarray}
Alternatively, the coupled equations can be rewritten as a single
vector equation involving only a n.n. interaction:
\begin{equation}
J(y,x)\Psi_{i+1}+J(x,y)\Psi_{i-1}+2h_i\Psi_i=E\Psi_i,
\end{equation}
where $J(x,y)$ is a $2 \times 2$ matrix whose nondiagonal entries are
zero while its diagonal elements are $J_y$ and $J_x$ (see Eqs. 4 and 5)
and $\Psi$ is a two-dimensional vector.

In the isotropic limit, the coupled equations become degenerate and the
two components of $\Psi$ become identical. The resulting equation is the famous
Harper equation\cite{Harper} which is a paradigm for quasiperiodic models
exhibiting the metal-insulator transition.
The anisotropic model describing a perturbed
Harper equation can also be written as
\begin{equation}
(H_h-2gH_g)(H_h+2gH_g)\Psi=E^2\Psi
\end{equation}
where $H_h$ denotes the Harper Hamiltonian consisting of the
first and the third term in Eq. (3) and $H_g$ denotes the part
of the Hamiltonian which exists only at a non-zero value of the
anisotropy $g$ defined as $J_x=1-g$ and $J_y=1+g$. We see that the anisotropy
perturbs the Harper squared model.

The fact that the anisotropy doubles the size of
the Hilbert space can be seen also in the semiclassical limit.
The anisotropic model
in the semiclassical limit describes the motion of a particle
with the Hamiltonian\cite{Wilkinson}
\begin{equation}
H = \frac{1}{2}(J_x+J_y)cos(p)+\frac{i}{2}(J_x-J_y)sin(p) +\lambda cos(x),
\end{equation}
where $p=\frac{\hbar}{i}\frac{d}{dx}$ is the momentum operator.
We see that the anisotropy perturbs the system
along a complex direction.
This explains why the Hilbert space
has doubled its dimension in the anisotropic case.

In the Ising limit $J_y=0$, the matrices $J(x,y)$ of the vector
equation are noninvertible. In this case, we will always work with
the TBM which contains only n.n. terms:
\begin{equation}
2J_xh_{i+1}\psi_{i+1} + 2J_xh_i\psi_{i-1} +(J_x^2+ 4h_i^2)\psi_i
=E^2\psi_i .
\end{equation}

An interesting limit of the Ising model is the case $J_x \rightarrow
\infty$ and $\lambda \rightarrow 0$\cite{Chaves} where the
TBM contains no diagonal disorder:
\begin{equation}
2h_{i+1} \psi_{i+1} + 2h_i \psi_{i-1} = E \psi_i
\end{equation}
It is easy to show that the model is always self-dual and has been numerically
shown to have only critical states. Therefore
we refer to it as the $critical$ model.
It should be noted that this TBM also describes the isotropic XY
model with $oscillatory$ $exchange$ interaction $J_x(i)=h_i$ and a
$constant$ field.\cite{Chaves} This should be contrasted with
the Harper model which describes the isotropic XY model with
$constant$ exchange and $oscillotory$ field.

The numerical phase diagram\cite{BR}  for the anisotropic
model describing a perturbed Harper model is shown in Fig. 1.
Whenever $\lambda$ and one of the
exchange interactions become equal either the E-C or C-L transition
takes place. The E-C transition corresponds always to a smaller
absolute value of $\lambda$, i.e. to the exchange interaction whose
absolute value is smaller. Whenever $J_x$ and $J_y$ differ
a fat $C$ phase is observed in the phase diagram. Therefore,
the fattening of the critical point is due to the breaking of
the $U(1)$ symmetry which is a consequence of the broken $O(2)$ symmetry in
spin space. An interesting consequence of the spectral and magnetic
interplay is the fact that
the onset of the C-L transition is coincident with the onset of
the magnetic transition to the long range order (LRO) where two-point
long range spin-spin correlations vanish.
These numerical
observations regarding the onset of E-C and C-L transitions
and the fact that C-L transition coincides with magnetic transition
still remain unexplained.

The energy spectrum of the anisotropic model is characterized by a gap
in the spectrum which vanishes along the conformal line. Therefore, unlike
the isotropic limit, $E=0$ is not an eigenvalue of the system in most
part of the phase diagram. As seen below,
this introduces another degree of
complexity in analysing the scaling properties of the model.

\section{ Fibonacci Decimation }

{\bf III.A Formulation }

One of the challenging questions posed by the fat C phase
is how the scaling properties vary within the phase. A priori there
are two main possibilities: either the scaling can be described by
only few different universality classes or the scaling properties
vary continuously through the phase. The former case would correspond
to a few distinct RG fixed points whereas the latter case could
be described by a line of RG fixed points in function space.

In recent years, various RG schemes have been proposed to study
the scaling properties of the critical point of the
Harper equation.\cite{Ostlund},\cite{rg},\cite{WRG},\cite{Jose}
Our methology is somewhat similar
to that of Ostlund et al.\cite{Ostlund} as it describes the scaling properties
of the wave functions for specific values of the energy. Furthermore, in
analogy
with their scheme, our decimation scheme has been applied to study the breakup
of KAM tori in the circle and standard maps.\cite{Ketoja},\cite{JK}
However, the method of Ostlund and Pandit\cite{Ostlund},
which was based on transfer matrices, had limited success due to the fact
that at the localization threshold
infinite products of transfer matrices diverged
for almost all values of the phase $\phi$. We propose a new decimation
scheme where instead of multiplying transfer matrices, the
TBM itself is decimated. The main advantage is the reduction
in the number of functions needed to carry out the renormalization.
The cost we have to pay is that our recursion relations will
be slighly more complicated. However, it turns out that with fewer
functions we are able to eliminate directions which lead to
divergences. This not only helps in approaching the RG problem but
also provides practical means of calculating various essential
quantities like the localization threshold and the minimum ($E_{min}$)
and maximum ($E_{max}$) energy eigenvalues
in general TBM's. The knowledge of machine precision $E_{min}$ and $E_{max}$
comes from the study of the states with Bloch index $k=0$ and $k=1/2$ in
addition to $k=1/4$ which respectively correspond to $E_{min}$, $E_{max}$
and $E=0$ quantum states. The previous studies of Ostlund and Pandit
\cite{Ostlund} and those of Dominguez, Wiecko and Jose\cite{Jose} were
restricted to the Bloch index $k=1/4$ corresponding to $E=0$ part of
the spectrum.

We will first describe the decimation scheme based on the Fibonacci
lattice sites.
This procedure describes only three critical scaling ratios
out of infinity that are required to fully characterize the wave
function.
However, the Fibonacci decimation scheme is quite sufficient to
confirm the fat C phase and to calculate critical parameters and
energies.
A decimation scheme based on an arbitrary path in the Farey tree
 (see sections 6 and 7)
will determine the infinite set of universal scaling ratios needed to
describe the universality completely. As discussed later in the paper,
the dominant peaks of the wave functions are described by $\sigma^3$
decimation.

In this scheme, all sites except those labelled by the Fibonacci numbers
$F_n$ ($n=0,1,2,...$) are decimated.
This results in a TBM connecting
a function $\Psi$ at two neighboring Fibonacci sites:
\begin{equation}
\Psi(i+F_{n+1}) = c_n(i) \Psi(i+F_n) + d_n(i) \Psi(i)
\end{equation}
In the Harper and the Ising limits, the above equation is a scalar
equation with scalar "decimation" functions $c_n(i)$ and $d_n(i)$.
For the general anisotropic case, $\Psi(i)$ is taken to be
the vector $(\psi_i ,\eta_i)$ of the
coupled eigenvalue equation and the multiplying factors $c_n$ and $d_n$
are $2 \times 2$ "decimation" matrices. In this case,
the name "decimation function" refers to  the entries of these matrices,
which are functions of the lattice site $i$.
The index $n$ labels the "level" of the decimation. As the fractional
part of $F_n \sigma$ decays to zero as $n \to \infty$, we expect
the scaling of the wave function for consecutive Fibonacci sites to show some
regularities which should also come up in the decimation functions.
Using the defining property of the Fibonacci numbers,
$F_{n+1} =F_n + F_{n-1}$, the following recursion relations
are obtained for $c_n$ and $d_n$\cite{Ketoja}:
\begin{eqnarray}
c_{n+1} (i)&=& c_n(i+F_n) c_{n-1} (i+F_n)-d^{-1}_n(i) d_{n+1} (i)\\
d_{n+1} (i)&=& -d_n (i)
[d_n (i+F_n )+\nonumber \\
& & c_n(i+F_n) d_{n-1} (i+F_n)] c_n^{-1} (i).
\end{eqnarray}
As seen from the above equations, the decimated matrices
describe a flow in Fibonacci space provided their inverse exist.

In order to iterate these equations, we need the initial conditions
for the matrices at the levels $n=1$ and $n=2$, say.
For the standard Fibonacci numbers with $F_1 =F_2 =1$, it is possible
to take $c_1$ to be the unity matrix and let $d_1$ vanish.
The non-trivial
initial conditions at $n=2$ are obtained directly from the
defining equations (4-5) for the TBM. The decimation could be
carried out also with other choices for $F_1$ and $F_2$ but
this would require more elaborate initial conditions.

Comparing our decimation scheme with that of Ostlund et al.\cite{Ostlund},
it should be noticed that our approach based on the special TBM form
requires half the number
of functions needed for the method based on transfer matrices. As discussed
below,
our new approach eliminates most of the divergences encountered in the previous
approach.

{\bf III.B Results and applications }

The recursion relations for the decimation functions were iterated
numerically for extremely large size systems (upto sizes 500,000)
required to study lattice of $F_{24}$ spins. The renormalized
$2 \times 2$ decimation matrices were then diagonalized.

Numerical iterations of the scalar decimation equations for the Harper and
the Ising models result in well-defined asymptotic solutions for the
functions $c_n$ and $d_n$ as $n\to \infty$ in both E and C phases.
For the vector decimation
of the general anisotropic model, asymptotic
decimation matrices
and their inverse were found to exist in the $E$ phase as well as
along the $E-C$ transition line in the phase diagram. However, along the
conformal line and within the fat C phase for the general anisotropic case
(except the Ising limit), the $c-$ or $d$-matrices
were non-invertible for some lattice sites
and hence our decimation equations could not be used there. We used
alternative means to study this part of the phase diagram as described in the
next section.

The asymptotic behaviour of the decimation functions as $n \to \infty$
is different in the E,C, and L phases and also depends sensitively on the
energy $E$ and the phase $\phi$.
In the L phase, $c_n$ always diverges and $d_n$ tends to zero
as one would expect by looking at Eq. (12). In the E and C phases,
the decimation functions remain bounded for all $n$ (except for
some cases in the E phase, see Table I).
Furthermore, at the band edges corresponding to the minimum and
maximum eigenvalue $E_{min}$ and $E_{max}$ (and also at the band
center for Harper), the decimation functions
converge on well-defined limit cycles provided the phase factor
$\phi$ is tuned to some special values. For arbitrary $\phi$ and also
for other eigenvalues, the decimation functions
oscillate in a rather irregular way with increasing $n$ converging perhaps
on a strange attractor.

Table I shows the trivial limit cycles for the E phase.
The asymptotic solutions are independent of the chosen value
of $\phi$ and hence do not depend upon the site index $i$.
This provides
a rather interesting example of the dimensional reduction for a system with
infinite degrees of freedom. It should be noted that the limit cycle is
almost independent of the anisotropy $g$.

For the Harper model and also along the conformal line in the phase
diagram the existence of a six-cycle can be seen for the zero-energy
state. For rest
of the phase diagram, the model has no zero-energy eigenstate
and hence the determination of a limit cycle requires eigenenergy to an
extremely high precision. The values of the other parameters determines
how  sensitive the solution is to an error
in the energy. Typically, in order to see the limit cycle with three digit
precision, energies are required to have machine precision.
In the worst case of a delayed crossover,
where the parameter values are
very close but not exactly on a transition line,
it may happen that machine precision energy may give an asymptotic
limit cycle with
single digit precision. Obtaining machine precision energies is almost an
impossible task
even for the tridiagonal matrices. (See Table
III)
Table II illustrates how the fixed point found in the E phase can be used
to numerically determine $E_{max}$ for an arbitrary
value of $\lambda$. A good initial estimate of $E_{max}$
can be obtained via a direct diagonalization of the TBM with
periodic boundary conditions.
This initial estimate can then be improved by imposing the fixed point
property. The resulting energy eigenvalues are
much more accurate than the ones obtained by the
diagonalization.\cite{footnote}

Unlike the trivial limit cycle of the E phase, the C phase is characterized by
non-trivial
asymptotic 6-cycles observed at the band edges
for carefully chosen values of $\phi$.
Furthermore, the
decimation functions depend upon $i$ (see Figs. 5 and 6).
 Our numerical results show that
$\phi=1/4$, for which the QP potential has the symmetry $h_{-i} = - h_i$,
always
gives a desired six-cycle throughout the fat C phase.
The value $\phi=0$ leads also to a reflection symmetry (due to
various symmetries in the anisotropic model the phase can be
defined mod $1/2$). The phase $\phi=1/4$ is however special because
for the part of the phase diagram where zero energy state is an
eigenstate (Harper model and the anisotropic model along the
conformal line), this value of $\phi$
causes the main peak of the wave function to lie
at $i=0$ and the resulting wave function is bounded. (See Section IV)

In the cases where $E=0$ is not an eigenstate,
$\phi=1/4$ was found to shift the main peak
in a very orderly fashion (related to number theory) for finite lattices.
For the Fibonacci lattices of size $F_k$, the main peak is located at $p_k$,
 where
$p_k$ satisfies the recursion $p_{k+1}-p_k = 4(p_k-p_{k-1})+(p_{k-1}-p_{k-2})$.
Therefore, the shifts in the peak positions $p_k-p_{k-1}$ give the rational
approximants of $\sigma^3$.
For the odd Fibonacci lattice sizes the shifts $p_k-p_{k-1}$,
are given by the even Fibonacci sequence $8,34,144...$.
while that for the lattices of size $8,34,144...$, the
shifts are $21,89,377,...$. The shift in the main peak between two
neighboring odd-even cases is $4, 17, 72, 305,...$. This systematic
shift in the main peak implies that the
continuously varying critical $\phi$ for locating the main peak at $i=0$ can
be found upto arbitrary precision by the formula
$\phi = \lim_{k \rightarrow \infty} <\sigma p_k+1/4>$, where $<\;\;>$
denotes the fractional part. For the fat C phase, this
phase factor is found to vary continuously as $p_0, p_1, p_2$ vary inside the
fat C phase. The knowlege of $\phi$ is essential in confirming the
universality as discussed in the next two sections.\cite{problem}

Detailed numerical iterations of the decimation equations reveal
that the fat C phase of the model
is described by $four$ distinct limit cycles:\\
(1) Spectral limit cycle along the E-C transition line.\\
(2) Conformal limit cycle along the conformal line (C-L line).\\
(3) Subconformal limit cycle, for the fat C regime bounded by
E-C and C-L transitions.\\
(4) Harper limit cycle corresponding to the isotropic limit of the model.\\

As stated earlier, all the four limit cycles were found for $E_{min}$ as well
as for $E_{max}$ (and also for E=0 for Harper). It is interesting that
the spectral limit cycle does not depend upon the
actual value of the $E_{min}$ and $E_{max}$ which varies continuously
along the E-C transition line. For the subconformal limit
cycle, both the energy and the strength of the field $\lambda$ were
irrelevant parameters. Therefore, the subconformal
limit cycle is identical to the limit cycle
observed for the critical model (Eq. 11).

Table III compares the eigenenergies obtained by diagonalizing
tri-diagonal matrices of varying sizes with those obtained by
the Newton method based on the existence of the six-cycle.
This table clearly shows the superiority of the decimation method compared
to the exact diagonalization even for tri-diagonal matrices.
As stated earlier, a very precise value of $E_{min}$ is
essential to confirm the C phase at subconformal points.
Near the conformally invariant point we observe that $E_{min}$
behaves like $J_x 2^z c(1-\lambda)^z$ with $c\approx 0.3708641926$ and
$z\approx 1.38897$.
This gives a rather good estimate for
$E_{min}$ through the whole fat C phase.
Close to the limit $\lambda=0$,
$E_{min}$ behaves like $J_x+E_{min}^c \lambda/2 +...$, where $E_{min}^c$
is the band edge of the critical model. This fact together with
Eq. (11) implies a continuous connection between the band edge of the critical
model and $E_{min}$ of the Ising model.

In order to confirm the universality of the observed limit cycles,
the decimation procedure was repeated for the
multiharmonic QP potentials
\begin{equation}
h(i)=\frac{\lambda}
{\sqrt{1+\alpha^2}}(cos(2\pi (i\sigma +\phi))+
\alpha cos(6\pi (i\sigma +\phi)))
\end{equation}
The observed limit cycles for this generic potential were identical to those
observed for a single harmonic field in all the universality classes discussed
above. Therefore, we conclude that
the fat C phase persists for generic
potentials.

Numerics to demonstrate the limit cycle in the decimation functions
for multi-harmonic field is complicated by the fact that
the phase boundaries have to be known to a very high precision.
We were again assisted by the existence of limit cycles
which provided a very efficient Newton method to locate the phase
boundaries.
In Table IV, we display the flow of the decimation functions to
an asymptotic 6-cycle
for the Ising model at the onset of the C-L transition where the model
is conformally invariant. The results of this table are used in determing
the C-L phase boundary as shown in table V.

The region of the phase diagram where the vector decimation cannot be applied
due to noninvertibilty of the decimation matrices,
will be studied in the
later part of the next section. Based on those studies, we conjecture that
these regimes do not introduce any new limit cycles.

\section{Symmetric and Asymmetric Wave Functions}

The existence of a $p$-cycle for the decimation functions often implies that
the self-similarity in the wave function is described
by the equation,\cite{Ostlund}
\begin{equation}
\Psi(i) \approx \Psi([\sigma^p i + 1/2])
\label{wf}
\end{equation}
($[\;\;]$ denotes the integer part).
This equality with $p=6$ (or $p=3$ if the absolute values are
considered)
was originally found for the Harper model at its critical
point. Our studies
generalizes the validity of this result for the
whole class of TBM's discussed here. In case where both
the decimation functions and the wave function are bounded,
a limit cycle for the decimations functions necessiates the above-form
self-similarity for the wave function and vice verse.\cite{div}

For a given eigenstate with known eigenvalue $E$, the
TBM provides an iterative scheme to obtain the wave function at all
sites in terms of $\Psi(0)$ and $\Psi(1)$. We choose $\psi_0=1$ and adjust
the phase factor $\phi$ so that the maximum of the absolute value
of the wave function $\psi$ is at $i=0$. This method always gives a bounded
solution for the wave function. The unknowns $\eta_0$, $\psi_1$,
and $\eta_1$ are then
determined requiring that the above equation becomes exact as $i$
tends to infinity. Eq. (12) implies
\begin{equation}
\Psi(F_{n+p})=C_{n,p} \Psi(F_n) + D_{n,p} \Psi(0)
\end{equation}
where
\begin{eqnarray}
C_{n,p} &=& c_{n+p-1}(0) c_{n+p-2} (0) ... c_n (0) \\
D_{n,p} &=& \sum_{i=0}^{p-1} C_{n+i+1,p-i-1} d_{n+i}(0)\;\;\;\;(C_{n+p,0}
= unity)
\end{eqnarray}
If there is an asymptotic $p$-cycle for the decimation functions,
$C_{n,p}$ and $D_{n,p}$ approach a fixed point in the limit
$n\to \infty$.
Assuming that $\Psi(F_{n+p})=\Psi(F_n)=\zeta \Psi(0)$ we can
write Eq. (17) as an eigenvalue equation and solve for the eigenvalue
$\zeta$ and the eigenvector $\Psi(0)$ (i.e. $\eta_0$). An estimate for
$\Psi(1)$
is then obtained by applying the recursion relation (12) backwards
$n+p-2$ times.
This section describes wave functions obtained in this way.

Figs. 2 and 3 show the wave functions corresponding to different parts of the
phase diagram.
It is interesting to note that for Harper, the wave function is symmetrical
about
$i=0$ wherease for the Ising case at the conformally invariant point,
it is completely asymmetrical. The symmetry in Harper is obvious from
the TBM.
The asymmetry of the $E=0$ eigenstate in the Ising case
is a consequence of the fact that in this limit the wave function
is given by the simple recursion $J_x \psi_{i-1} +2h_i \psi_i =0$.
Because $h_0 =0$ with $\phi=1/4$, we obtain $\psi_i =0$ for all $i<0$.
A more physical picture of the wave function could be obtained by
choosing another phase factor for which $h_i$ would be nonvanishing for
all finite $i$. However, even in this case there would always exist
an $i$ so that $h_i \approx 0$ and the wave function
would essentially vanish on a finite left-hand-side neighborhood of
that lattice site.
The asymmetry
can be understood also from the semiclassical
analysis of the TBM.\cite{Wilkinson}
In the Ising case, the eigenvalue equation
 in the semiclassical limit is (from Eq. 9)
\begin{eqnarray}
exp(-ip) \psi(x)+ \lambda cos(x) \psi(x) &=& E \eta(x)\\
exp(ip) \eta(x)+ \lambda cos(x) \eta(x) &=& E \psi(x)
\end{eqnarray}
For $E=0$, if $cos(x_0)=0$ for some $x_0$, then
$exp(-ip) \psi(x_0) = exp(ip) \eta(x_0) = 0$.
Since $p$ is the generator of space translation,
the above equation implies that $\psi(x)$ and $\eta(x)$ respectively
 vanish to the left and right of $x_0$.

The figures suggest that the phase diagram for arbitrary $g$
can be classified as
either symmetric or asymmetric. Along the conformal
line in the phase diagram, the wave function is always
asymmetric about the center. However, in the subconformal region
of the C phase
the system slowly
recovers its symmetry asymptotically. For the critical model,
the wave function fully restores its symmetry
 and in fact
qualitatively resembles the Harper case. For the critical model, the phase
$\phi =(3-2\sigma )/4$ setting
the main peak at $i=0$ is such that the potential
has the symmetry $h_{-i} = - h_{i+1}$. This implies the reflection
symmetry for the wave function. Furthermore, the E-C transition
is also characterized by an asymptotically symmetric wave function.
Fig. 4 shows the Hull function, obtained by plotting the wave function
$|\psi_i|$ as a function of $<i\sigma>$, in the four universality
classes.

In the region where the vector decimation could not be implemented, we have
to use other methods.
Along the conformal line for $E=0$ state, the coupled equations
(4-5) decouple. It turns out that either the forward or backward
iteration of Eq. (4) is extremely stable beginning with an arbitrary
initial condition. In other words, we can give $\psi_{\pm N}$ an
arbitrary initial value and iterate in the stable direction to obtain
$\psi_i$ for $i=...-M,-M+1,...,M$. These $\psi$-values are very accurate
if $N$ is much larger than $M$. In the subconformal region of the
anisotropic model, the wave function is obtained by the
diagonalization method and thus it is there not so accurate as in other
parts of the phase diagram.

In the next section, we provide a quantitative characterization
of the universality classes.

\section{Infinite Set of Universal Scaling Ratios}

Ostlund and Pandit\cite{Ostlund} described the universality
of the Harper critical point in terms of the scaling ratio
\begin{equation}
\zeta = \lim_{n \rightarrow \infty} |\psi(F_{3n})/\psi(0)|,
\end{equation}
This scaling ratio along with Eq. (16)
clearly describes the wave function in the C phase
by implying that it is neither extended nor localized. Eq. (22)
describes the decay of wave functions at asymptotic Fibonacci
sites with respect to the central peak. The existence of a limit cycle with
period $3$ in the decimation equation (considering the absolute
values) gives three different values
of $\zeta$. Two of them are listed in Table VI. The interesting question
is what happens to the similar scaling ratios at sites other than
the Fibonacci's. Does the wave function
repeat itself in the same way as it does for the Fibonacci sites?
Our detailed numerical studies suggest that it is true for all sites
that a critical wave function repeats itself:
i.e for $every$ given site, there exists a whole
sequence of sites $Q_k$ where the wave function approaches
the same amplitude which is a universal fraction of the main peak. This
sequence is given by the recursion relation
\begin{equation}
Q_{k+1}=4Q_k+Q_{k-1}.
\end{equation}
The recursion is such that $Q_k /Q_{k+1} \to \sigma^3$
as $k\to \infty$. This implies that
the scale invariance of the wave function
is described in terms of an infinite
set of scaling ratios $\zeta$,
\begin{equation}
\zeta(Q_0 ,Q_1 ) = \lim_{k \rightarrow \infty} |\psi(Q_k)/\psi(0)|,
\end{equation}
where $Q_k$ is obtained by the above recursion relation.
The set obtained by varying the two integers $Q_0 ,Q_1$
is complete and hence specifies the location and
height of all the peaks of the wave functions.
Stated differently, this result implies that the wave function
at every $Q_k$ site
repeats itself at sites given by the above recursion
relation. It is easy to see that this
generalization includes the previous results\cite{Ostlund} related
to the Fibonacci sites as a special case. (See Table VI)

We next investigate the sequences that result in the $dominant$ peaks
in the wave function. Our detailed studies show that
out of infinity of the scaling ratios, the $\zeta$'s for the
dominant peaks are obtained from those values of $Q_0$ and $Q_1$ that are of
the order of $4$.

Furthermore, it turns out that the dominant peaks can be labeled
in two ways. This also helps in regrouping the infinite set of $\zeta$'
s. First, with every sequence determined by $Q_0$ and $Q_1$,
we associate a
set of $n$ "harmonic" sequences ($nQ_0,nQ_1, ...$).
each corresponding to its own unique $\zeta$. We notice
 that the dominant
$\zeta$'s correspond to lower harmonics as shown in Table VII.
It is interesting to note that the ($0,2,8,34,...$) sequence associated
with the Fibonacci numbers can be considered as the second harmonic of
the sequence ($0,1,4,17,72,...$) associated with the rational
approximants of $\sigma^3$.
It should be noted that unlike the other two sequences ($1,3,13,55...$) and
($5, 21, 89.....$), the sequence consisting of even integers ($0,2,8,34,...$)
does not belong to the sequences associated with $\sigma^3$. However, the fact
that it can be regarded as a higher harmonic of one of the $\sigma^3$
sequence suggests that the important peaks are determined by $\sigma^3$
periodicity and its harmonics, instead of $\sigma$ periodicity.

Alternatively,
the location of the dominant peaks in the wave functions can also be related
to the number theoretical properties of $\sigma^3$ using the Farey tree
\cite{Farey}.
Denoting the left and right branches of the Farey tree by $0$ and $1$,
the rational approximants of $\sigma^3$ can be represented by the path
$000111100001111...$ which we will denote as $0^3 1^4 0^4 1^4 ...$.
We conjecture that the dominant peaks in the wave functions occur
on the sites which correspond to the denominators of the rational numbers
on this main path and on the side paths that differ
minimally from the main path.

\section{ Decimation Scheme Based on Arbitrary Farey Paths }

The infinite set of scaling rations can be obtained by generalizing
the Fibonacci decimation scheme. As described in the previous
section, every site in the spin chain can be labelled by a
symbol sequence corresponding to a definate path in the Farey tree.
We develop a decimation scheme based on this.
The basic idea underlying the decimation can be understood by considering three
Farey levels denoted as $f$ (father), $m$ (mother), $d$ (daughter).
Our decimation equations are,
\begin{eqnarray}
\Psi(i+Q_d) &=& c_n(i) \Psi(i+Q_m) + d_n(i) \Psi(i)\\
\Psi(i+Q_d) &=& a_n(i) \Psi(i+Q_f) + b_n(i) \Psi(i)
\end{eqnarray}
where $Q_d=Q_m+Q_f$.
In order to obtain the recursion relations for the decimation functions
$c_n,d_n,a_n,b_n$, we consider
the next level denoted as
$g$ (granddaughter). Now, $Q_g=Q_f+Q_d$ or $Q_g=Q_m+Q_d$
depending upon whether the $Q_g$ is obtained by taking a step in the same
direction or the different direction from the one from $Q_m$ to $Q_d$.
The above two routes to $g$ level can be denoted as $0(1)
\rightarrow 0(1)$ and $0(1) \rightarrow 1(0)$.

 The recursion relations
for $0(1) \rightarrow 0(1)$ are
\begin{eqnarray}
a_{n+1}(i)& =& c_n(i+Q_f)a_n(i)+d_n(i+Q_f)\\
b_{n+1}(i) & = & c_n(i+Q_f)b_n(i) \\
c_{n+1}(i) & = & c_n(i+Q_f)+d_n(i+Q_f) a_n^{-1} (i)\\
d_{n+1}(i) & = & -d_n(i+Q_f)b_n(i) a_n^{-1} (i)
\end{eqnarray}
and for $0(1) \rightarrow 1(0) $
\begin{eqnarray}
a_{n+1}(i)& =& a_n(i+Q_m)c_n(i)+b_n(i+Q_m)\\
b_{n+1}(i) & = & a_n(i+Q_m)d_n(i)\\
c_{n+1}(i) & = & a_n(i+Q_m)+b_n(i+Q_m) c_n^{-1} (i)\\
d_{n+1}(i) & = & -b_n(i+Q_m)d_n(i) c_n^{-1} (i)
\end{eqnarray}
It should be noted that the Fibonacci decimation corresponds to
the symbol sequence $101010...$ in which case the recursion relations
can be written in the form in which $a_n$ and $b_n$
become redundant.

In order to explain the infinite set of $\zeta$'s we first need
the decimation functions along the "main" Farey route to
$\sigma^3$  which defines
four universal scaling ratios (there is an 8-cycle in the decimation
functions). Other universal $\zeta$'s are then obtained by
determining the decimation functions along finite side paths
from the main route. It should be noted that the most dominant peak in the
wave functions (corresponding to largest $\zeta$) occurs at $1,4,17,72,...$
(see Figs. 2 and 3) and can be obtained only by $\sigma^3$ decimation.
In the previous RG scheme\cite{Ostlund}
the scaling ratio $\zeta$ corresponding to this sequence was not
predicted.

\section{Renormalization Group Equation}

The universal functions and the
scaling ratios can in principle be obtained from RG equations.
We have already derived recursion relations for the decimation
functions or matrices which can be used as the starting point for the
definition of a suitable renormalization operator.
However, in order to get the final equations in a solvable
form it is essential to get rid of the discrete lattice index $i$.
Because the QP potential has been constructed using a period-1
function, the decimation functions/matrices $a_n (i), b_n(i),c_n(i),
d_n(i)$ can
be thought of being functions of the fractional part of
$i\sigma$, denoted by $<i\sigma>$, only. This gives us a continuous
variable. But we also need to consider the scaling.
Assuming a $p$-cycle for the decimation functions at $i=0$,
we observe that if $a_n, b_n, c_n, d_n$ are defined
as functions of the renormalized variable
$x=(-\sigma)^{-n} <i\sigma>$, any decimation
function of the level $n$ maps roughly onto the corresponding function
of the level $n+p$ for $x\in [0,1]$ (see Figs. 5 and 6). This indicates
that $x$ provides us with a proper continuous variable.

The next step is to rewrite the recursion equations for the
decimation matrices/functions in a form which uses the above
renormalized variable $x$.
For simplicity,
we restrict ourselves to the Fibonacci decimation.
By using the relation $F_n \sigma = F_{n-1} -(-\sigma)^n$,
the nonlocal recursion relations (13-14) can be transformed into
local RG equations
\begin{eqnarray}
c_{n+1} (x)&=& c_n(-\sigma x-1) c_{n-1} (\sigma^2 x+\sigma)\nonumber \\
& & -d_n^{-1}(-\sigma x)d_{n+1}(x)\\
d_{n+1} (x)&=& - d_n (-\sigma x)
[d_n (-\sigma x-1) +\nonumber \\
& & c_n(-\sigma x-1) d_{n-1} (\sigma^2 x+\sigma)] c_n^{-1} (-\sigma x)
\end{eqnarray}
This defines our renormalization operator which can be now applied
to determining the universal cycles. We tried that in the case
of the critical Harper model ($E=0$). Our detailed
numerical studies showed that
$(c_n (x), d_n (x))$ approached a six-cycle
$(c^*_1 ,d^*_1 )\rightarrow (c^*_2, d^*_2)\rightarrow
(c^*_3 ,d^*_3 )\rightarrow (-c^*_1, -d^*_1)\rightarrow
(c^*_2 ,-d^*_2 )\rightarrow (-c^*_3, d^*_3)\rightarrow
(c^*_1 ,d^*_1)$ where all the functions of the six-cycle appeared
bounded and smooth for the full messure of $x$ values. This suggested
the functions $c^*_n$, $d^*_n$ could be expanded
in a power series and one could use
the Newton method to solve for the coefficients.
However, simple circular complex domains centered on the real axis
did not give convergent results. We believe
that the equations can be solved using more elaborate domains.
It is probably necessary
to let the centers of at least some of the complex disks to be located
off the real axis so that the radius of convergence can be increased.
This appears true especially for the expansion of the inverse of $c_n$
for which this radius is only unity if the center of the corresponding
domain lies at the origin of the complex plain. Therefore, the determination
of scaling properties using eigenvalues of linearized RG equations
about the limit cycle remains open.

As to the RG for an arbitrary Farey path, the main difficulty
arises from the number-theoretical problem of generalizing
the relation giving the fractional part of $F_n \sigma$ to
products of the type $Q_f \sigma$ and $Q_m \sigma$ where
$Q_f$ and $Q_m$ are arbitrary integers appearing in the Farey tree.
All we can prove at the moment is that with the recursion
(23) the fractional part of $Q_k \sigma$ (with arbitrary
$Q_0,Q_1$)  approaches either
$0$ or $1/2$ and the rate of convergence is given by
$(-\sigma)^3$. However, there are still a lot of details
which need to be worked out in the general case.

\section{ Summary of the Universality Classes }
Based on all the infinite set of scaling ratios, we conjecture that
the C phase of the model is governed by four different fixed points
of the renormalization group defining four universality classes,
which we refer as spectral, conformal, subconformal and Harper.
We show that the
anisotropy is a relevant parameter as
an infinitesimal anisotropy takes the system to one of the three possible
new universality classes, away
from the Harper universality describing the isotropic limit of the model.
Except for the conformal universality, the other three universality
classes exhibit an asymptotic reflection symmetry of the wave function.

We checked in all the above four cases that the universality
is rigid with respect to replacing the
single harmonic $h_i$ by an odd two-harmonic function.
Both the infinite set of $\zeta$'s
as well as the decimation functions were found to be universal.

We also computed the exponent $\beta$ defined by Ostlund et al.
\cite{Ostlund}
 which measures the degree of localization around the
site where the wave function is peaked.
\begin{equation}
-\beta = \frac{1}{p ln(\sigma^{-1})}\lim_{n \rightarrow \infty} [ ln[
\frac{L_n}{L_{n+p}}]]
\end{equation}
where
\begin{equation}
L_n = \frac{1}{Q_n}\sum_{i=-Q_n}^{Q_n} \psi_n^2
\end{equation}

The $p$ in the above equation is the period of the limit cycle.
Unlike the scaling ratio $\zeta$, no close form expression for
$\beta$ could be found in terms of the decimation functions.
$\beta$ was computed directly from the wave function.

The exponent $\beta$ is $0$ in the E phase, $-1$ in the L phase and
varies between these two values in the C phase.
Table VIII shows $\beta$ in four different universality classes.
As expected the absolute value of $\beta$ is greater at the
conformal fixed point compared to the subconformal case. Comparing
$\beta$ at the onset of localization in the Harper and Ising-like cases
leads us to
conclude that the LRO slightly suppresses the degree of localization.
However, the degree of localization remains constant inside the C phase.
Since the subconformal C phase existing in finite window in $\lambda$
is described by a single fixed point this implies that $\lambda$ is an
irrelevant
parameter in this phase.

All the other universality classes except the spectral one can
be described by scalar decimation functions. This made the study of the
spectral
universality a lot more difficult. Additional complications came from
the facts
that $E_{min}$ was non-zero and the phase factor $\phi$ was continuously
changing. This explains why the wave function and the scaling
ratios $\zeta$ are known with less precision in that case than for the
the other three universality classes.

It should be noted that the universality classes obtained here
are for the fractal
quasi-particle {\it wavefunctions} and do not tell us anything about
the scaling properties of the fractal quasi-particle energy
spectrum. Very recently, Chaves\cite{jcc} showed that the energy level
statistics follow inverse power law distribution with the exponent $3/2$
throughout the fat C phase. This implies that the fractal dimension of
the C phase is a constant throughout the phase and
 is equal to $.5$.\cite{Geisel}. However, the multifractal
analysis of the spectrum\cite{BR} showed that the $f(\alpha)$ curve\cite{BR}
describing the scaling properties of the energy spectrum
is different in the four universality
classes. Therefore,
we conjecture that even
the multifractals describing the quasi-particle energy
fall within the same four classes. We believe that the exact RG treatment
describing the spectrum as a whole as
proposed by Wilkinson\cite{WRG} could be generalized to confirm this
conjecture.

\section{Conclusions}
The work described here confirms that the anisotropic spin model describing
the perturbed Harper with broken $U(1)$ symmetry
fattens the critical point of the Harper model. In analogy with the mode-locked
phase, this fat C phase
existing in a finite window
can be described as $C-locked$ phase whose scaling properties fall within
four different universality classes.
The conformal and the spectral
universality classes define the boundaries of the phase diagram which
sandwich the subconformal class. In the isotropic limit, the three
universality classes degenerate to the Harper universality.
It is rather interesting that the subconformal and conformal parts
of the phase diagram belong to different universality classes.
This provides a novel example of spectral and magnetic interplay
showing how a magnetic phase transition (which occurs also in a
periodic model) affects the critical phase characteristics of
aperiodic systems. The spectral transition which exists only in an
aperiodic system (and does not affect the magnetic properties such as the
long-range correlation function) defines its own universality class
distinct from the conformal and the subconformal cases.

The E phase of the quasiperiodic models is the KAM phase where the
wave functions are extended and hence the quasiperiodicity has minimal
effects. As shown in the Fig. 7,
in contrast with the C phase, the Hull function is smooth here.
It is rather surprising that the universality
in this phase is almost independent of the anisotropy $g$.
For the periodic system, the anisotropy is a relevant parameter as it
provides a preferred axis for spin alignment. The fact that the
universal functions
of the E phase are insensitive to the anisotropy adds a
new dimension to the KAM theory: the smooth Hull function of
the integrable limit $\lambda=0$ survives not only the $\lambda$
perturbation in the E phase, but the anisotropy
leaves the universaliry class unaltered.
This is different from the C phase where an infinitesimal anisotropy alters the
universal characteristics. Therefore, we conclude that the anisotropy
responsible for the long range
order does not affect the KAM phase and the quasiperiodicity responsible
for the spectral transition does not affect the long range order.

It is interesting to compare our results with those obtained
with a TBM related to the stability of the ground state in
the classical Frenkel-Kontorova model at the onset of pinning-depinning
transition.\cite{JK} This pinning-depinning transition
corresponds to the breakup of KAM tori in area preserving maps.\cite{Mackay}
In this case the analog of the fractal wave function in the C phase is
the derivative of
the Hull function describing a QP critical ground state (Fig. 8).
The reason why here the wave function corresponds to the
derivative of the Hull function rather than the Hull function itself
is the fact that the TBM follows from a variation of the ground state.
With the initial conditions on the dominant symmetry line (this
is analogous to our choice of the phase $\phi$), the Fibonacci decimation
leads to an asymptotic universal fixed point.\cite{Ketoja}
It is interesting to speculate the existence of a C-locked critical
Hull function in some generalized classical Frenkel-Kontorova models
or the corresponding area preserving maps.\cite{Mackay}

Very recently, Faddeev et al.\cite{BA} have shown that the Harper equation
describing isotropic limit of the XY model can be solved using generalized
Bethe Ansatz.
It will be interesting to know if the coupled equations (4-5) describing the
anisotropic model are amendable to similar treatment. Since in the
case of a periodic system ($\sigma=0$), the anisotropy introduces an integrable
perturbation on the isotropic model, it is possible that even with
quasiperiodicity that may be the case. We hope that our studies
will motivate rigorous analysis of the quantum spin models discussed
here.

\acknowledgements

The research of IIS is supported by a grant from National Science
Foundation DMR~093296. JAK is grateful for the hospitality
during his visit to the George Mason University where this work was
initiated. IIS would like to acknowledge the hospitality
of Research Institute for Theoretical Physics at Helsinki
during the course of this work.

\begin{figure}
\caption{ Phase diagram for the quasi-particle excitations:
With $J_x=1-g$ and $J_y=1+g$, the figure shows the E, C and
L phases in $\lambda-g$ plane. In the E phase, wave functions are
extended and hence resemble those of a periodic system. This phase
is also called the KAM phase as the Hull functions are smooth in this case.
In the C phase the wave functions are self-similar (see Figs. 2 and 3)
exhibiting atmost algebraic localization while in the L phase, they are
exponentially localized. The onsets of E-C
and C-L transitions are respectively the spectral line  (thin line) and
the conformal line (thick line). The two lines meet in the
isotropic (Harper) limit. The $g=-1$ describes the Ising limit with no E or
KAM type phase.}
\label{fig1}
\end{figure}

\begin{figure}
\caption{(a-e) show the absolute values of the wave functions, along
the line in the phase diagram where the model is conformally
invariant ($E_{min}=0$). For fixed $J_x=1>J_y \geq 0$, the conformal line
is obtained by setting $\lambda=1$ and varying $J_y$ (see ref. 5):
(a) $J_y=0$ , (b) $J_y=.99$, (c) $J_y=.999$, (d) $J_y=.9999$, and
(e)$J_x=J_y=1$ . The figures (a) and (e) correspond respectively to
the Ising and isotropic limits while (b),(c) and (d) are Ising like.
We see that the wave functions are highly asymmetric about
the peaks in the Ising and Ising like cases. However, in
the isotropic limit, the wave functions recover
the symmetry.}
\label{fig2}
\end{figure}

\begin{figure}
\caption{(a) shows the wave function $below$ the conformally invariant
point for the Ising model ($J_y=0$) with $\lambda=.95$.
(b) is the wave function of the critical model with no
diagonal disorder. (c) shows the wave function at the onset of the
spectral transition with $J_x=1$, $J_y=1.5$, and $\lambda=1$.
In all these cases, the wave function is always
asymptotically symmetrical (about $i=0$). In the limit
$\lambda \to 0 $, but finite, the wave function is
fully symmetric (b).}
\label{fig3}
\end{figure}

\begin{figure}
\caption{(a-d) respectively show the Hull function $|\psi(x) |$ vs
$x=<i\sigma>$
in the four universality
classes, namely the Harper $(E=0)$, conformal ($J_y=0$, $J_x=1$,
$\lambda=1$, $E=0$),
subconformal (critical model with $\lambda=.5$, $E_{min}= -1.420111920492815$),
and spectral ($J_x=1$, $J_y =1.5$,
$\lambda=1$, $E_{min}=0.17429863452742$).}
\label{fig4}
\end{figure}

\begin{figure}
\caption{(a-d) show the decimation function
$c_n (x)$ ($n=6-9$) for the
critical Harper model with $E=0,\phi=1/4$. As a
consequence of the 3-cycle for the absolute values of the decimation function,
$c_6(x)= \pm c_9(x)$. }
\label{fig5}
\end{figure}

\begin{figure}
\caption{(a-d) show the decimation function
$c_n (x)$ ($n=6-9$) for the
Ising model with $E=0,\phi=1/4$ at the conformally invariant point. As a
consequence of the 3-cycle for the absolute values of the decimation function,
$c_6(x)= \pm c_9(x)$. }
\label{fig6}
\end{figure}

\begin{figure}
\caption{
The wave function (a) and the smooth Hull function (b)
in the E phase at $\lambda=.5$, $J_x=1$, $J_y=1.5$,
and
$E=E_{max}= 2.620638044392$. Unlike the C phase, the Hull function is
smooth.}
\label{fig7}
\end{figure}

\begin{figure}
\caption{The derivative of the Hull function
at the onset of breaking of analyticity in the Frenkel-Kontorova model.
}
\label{fig8}
\end{figure}

\begin{table}
\caption{ The limiting subcritical (E phase) decimation functions for
the general anisotropic model ($g\not= 0$) and the Harper limit ($g=0$).
As the functions do not depend on the
phase $\phi$, there cannot be any dependence on the lattice
site $i$ either. For the general anisotropic case, the
cycles are seen only after diagonalizing the decimation matrices.
For the Harper limit $E_{min} =-E_{max}$ while for the
anisotropic model $E_{min}$ is positive. Note that $E_{min}$ of the anisotropic
case goes into $E=0$ of the Harper model as $g$ goes to zero. }
\begin{tabular}{cccccc}
 & & & & &\\
$E$ &  $c^*$  & & & $d^*$ & cycle   \\
 & & & & &\\
\tableline
 & & & & &\\
$g\not=0:$ & & & & &\\
 & & & & &\\
$E_{min}$ & $1,0,\infty$ & & & $0,1,\infty$ & 3\\
 & & & & &\\
$E_{max}$ & $\sigma^{-1}$ & & &$-\sigma$ & 1   \\
 & & & & &\\
$g=0:$ & & & & &\\
 & & & & &\\
$0$ & $1,0,-\infty,-1,0,-\infty$ & & & $0,-1,-\infty,0,1,
\infty$ & 6 \\
 & & & & &\\
$E_{min}$ & $\sigma^{-1}, -\sigma^{-1}, -\sigma^{-1}$ & & &
$\sigma,-\sigma,\sigma$& 3 \\
 & & & & &\\
$E_{max}$ & $\sigma^{-1}$ & & & $-\sigma$ & 1 \\
 & & & & &\\
\end{tabular}
\label{table1}
\end{table}

\begin{table}
\caption{ Estimates for $E_{max}$
of the Harper model with $\lambda =.5$ obtained from the
condition $c_n (0) =\sigma^{-1}$. The $n$ here refers to the $n^{th}$ order
Fibonacci number with $F_{25}=46368$.}
\begin{tabular}{cccccc}
 & & & & &\\
 & &$n$  &  $E$ & &   \\
 & & & & &\\
\tableline
 & & & & &\\
& &10 & 2.144122230954376 & &\\
 & & & & &\\
& &13 & 2.144103497103577  & &\\
 & & & & &\\
& &16 & 2.144103742043903 & &\\
& & & & &\\
& &19 & 2.144103738794523 & &\\
 & & & & &\\
& &22 & 2.144103738837256 & &\\
 & & & & &\\
& &25 & 2.144103738836694 & &\\
 & & & & &\\
\end{tabular}
\label{table2}
\end{table}

\begin{table}
\caption{$E_{min}$ for the Ising model obtained
by the diagonalization (size $N$ finite)
and the decimation scheme ($N=\infty$). }
\begin{tabular}{cccccc}
 & & & & &\\
$\lambda$ & N & $E_{min}$ \\
 & & & & &\\
\tableline
 & & & & &\\
0.95 & 1597 & 1.51567E-2\\
    & 2584 & 1.51484E-2\\
    & 4181 & 1.51481E-2\\
    & $\infty$ & 1.514806760988E-2 $\pm$ 1.E-14\\
0.995 & 4181 & 6.1929E-4\\
     & 6765 & 6.1862E-4\\
     & 10946 & 6.1853E-4\\
     & $\infty$ & 6.18420285E-4 $\pm$ 1.E-12\\
 & & & & &\\
\end{tabular}
\label{table3}
\end{table}

\begin{table}
\caption{ The decimation functions at $i=0$
for the Ising model with $\lambda=1$, $E=0$, $\phi=1/4$. The table clearly
shows the asymptotic six-cycle. }
\begin{tabular}{cccccc}
 & & & & &\\
$n$ &  $c_n (0)$  & $d_n (0)$  \\
 & & & & &\\
\tableline
 & & & & &\\
           3 & -2.245709795011781 &    -0.6002603574022150 \\
           4 &  8.109008292790231 &     -1.272168729380112 \\
           5 &  1.010919961015582 &    -0.3666408658531275 \\
           6 &  3.867229946686237 &    -0.7949593051332248 \\
           7 & -7.553298439810634 &      1.182294059162462 \\
           8 &  1.100178678647771 &     0.3764230488289012 \\
           9 & -3.710289833700136 &    -0.7701746581532440 \\
          10 &  7.758396265078099 &     -1.194676151297624 \\
          11 &  1.085391362405107 &    -0.3738057364522718 \\
          12 &  3.746636404387838 &    -0.7734760239494146 \\
          13 & -7.718184336265150 &      1.192015324586235 \\
          14 &  1.089002996132289 &     0.3742871587654810 \\
          15 & -3.738988672892117 &    -0.7728088313940935 \\
          16 &  7.727825304953185 &     -1.192588926144134 \\
          17 &  1.088173364711348 &    -0.3741752202327470 \\
          18 &  3.740771704903936 &    -0.7729526819664260 \\
          19 & -7.725577268094126 &      1.192454968911469 \\
          20 &  1.088369359725458 &     0.3742010466453257 \\
          21 & -3.740355675780264 &    -0.7729194653703071 \\
          22 &  7.726105073162105 &     -1.192485868259755 \\
          23 &  1.088319511696401 &    -0.3741951424798995 \\
          24 &  3.740445328629009 &    -0.7729293725799985 \\
          25 & -7.726061715596284 &      1.192499566226364 \\
 & & & & &\\
\end{tabular}
\label{table4}
\end{table}

\begin{table}
\caption{ Estimates for the conformally invariant point of
the Ising model where the third harmonic
with $\alpha=.2$
has been added to $h(i)$. The estimates were obtained  by imposing the
condition $c_n (0) =1.08832$ for $\phi=1/4$
(see the first column corresponding to $n=23$
in Table IV). }
\begin{tabular}{cccccc}
 & & & & &\\
 & &$n$  &  $\lambda$  & &  \\
 & & & & &\\
\tableline
 & & & & &\\
 & &11 & 1.367474785 & & \\
 & & & & & \\
 & &14 & 1.366202286 & &\\
 & & & & &\\
 & &17 & 1.366282249 & &\\
 & & & & &\\
 & &20 & 1.366277923 & &\\
 & & & & &\\
 & &23 & 1.366278173 & &\\
 & & & & &\\
 & &26 & 1.366278174 & &\\
 & & & & &\\
\end{tabular}
\label{table5}
\end{table}

\begin{table}
\caption{$\zeta$'s corresponding to some of the dominant peaks in the
Harper, Ising (both conformal and subconformal), and the spectral
universality classes. }
\begin{tabular}{cccccc}
 & & & & &\\
 sequence &  Harper & conformal & subconformal & spectral\\
 & & & & & \\
\tableline
 & & & & & \\
0,1,4,17,72,... &  .516 & .744 & .676 & .72\\
 & & & & & \\
1,1,5,21,89,... &  .315   &  .622 & .590 & .54 \\
 & & & & &\\
 1,2,9,38,... &  .256 & .297 & .337  & .58 \\
 & & & & &\\
1,3,13,55,... &  .143 & .496 & .231 & .37 \\
 & & & & &\\
\end{tabular}
\label{table6}
\end{table}

\begin{table}
\caption{ $\zeta$'s corresponding to a sequence and its harmonics for
the Harper model. }
\begin{tabular}{cccccc}
 & & & & &\\
harmonic & sequence & $\zeta$\\
 & & & & &\\
\tableline
 & & & & &\\
1 & 3, 13, 55, 233,... & 0.143\\
2 & 6, 26, 110, 466,... & 0.1456\\
3 & 9, 39, 165, 699,... & 0.1485\\
4 & 12, 52, 220, 932,... & 0.0183\\
5 & 15, 65, 275, 1165,... & 0.025\\
 & & & & &\\
\end{tabular}
\label{table7}
\end{table}

\begin{table}
\caption{ The exponent $\beta$ for the four universality classes. }
\begin{tabular}{cccccc}
 & & & & &\\
Universality Class & $\beta$\\
 & & & & &\\
\tableline
 & & & & &\\
Harper & -.639 $\pm$ .005\\
conformal & -.602 $\pm$ .001\\
subconformal & -.423 $\pm$ .001\\
spectral & -.175 $\pm$ .005 \\                     \\
 & & & & &\\
\end{tabular}
\label{table8}
\end{table}

\begin{references}
\bibitem{footnote} We find the same 3-cycle as for $E_{min}$ in the
Harper model also  just below $E_{max}$
at $E=1.943387097623$ with $\lambda=.5$. The significance of
this cycle is not clear to us.
\bibitem{problem} For reasons which we belive to be related to the asymptotic
validity of the equation for determining $\phi$, this method of determining
the phase factor was not always succesful.
\bibitem{div}
Our numerical studies suggest that it is possible that
a) the wave function has the self-similarity but the
asymptotic decimation functions are ill-defined,
b) the decimation functions show a $p$-cycle but
the wave function is unbounded (e.g. in subconformal region of the
Ising model with $\phi=1/4$).
\end{references}
\end{document}